\documentstyle[preprint,aps,epsf,eqsecnum]{revtex}
\begin{document}
\preprint{\vbox{\hbox{BARI-TH/97-250 \hfill} 
                \hbox{May 1997 \hfill}
                \hbox{Revised version \hfill} }}
\title{QCD Interactions of Heavy Mesons with Pions \\
by Light-Cone Sum Rules}
\author{P. Colangelo $^1$ and  F. De Fazio $^{1,2}$}
\address{
$^1$ I.N.F.N., Sezione di Bari, Italy\\
$^2$ Dipartimento di Fisica, Universit\'a di Bari, Italy}

\maketitle

\begin{abstract}
Light-cone QCD sum rules are employed to compute the strong coupling constants: 
$g_{B^* B^* \pi}$, $g_{B_1 B_0 \pi}$ and $g_{B_1 B^* \pi}$,  where 
$(B,B^*)$ and $(B_0,B_1)$ are negative and positive parity $(0^-,1^-)$ and
$(0^+,1^+)$ $\bar q Q$ doublets. 
The couplings are calculated both for finite values of the heavy quark mass
and in the infinite heavy quark mass limit, 
deriving sum rules for $m_Q \to \infty$. 
\\
\end{abstract}
\vspace*{1cm}
\newpage
\section{Introduction}

The recognition of the approximate symmetries of QCD, 
explicitely or spontaneously broken,
represents a basic step for the understanding and the description
of the strong interactions in processes involving hadrons. 
A remarkable example, in the sector of systems containing one heavy quark, is 
the $SU(2 N_f)$ heavy flavour-spin symmetry
holding in the infinite heavy quark mass limit: 
$m_{Q_f} \to \infty$. In this limit, 
the masses of the $N_f$ heavy flavoured quarks are irrelevant in the 
interactions with the hadronic light degrees of freedom, since
the heavy quarks only act as static colour sources.
Moreover, the decoupling of the gluon from the spin $s_Q$ of the heavy quark
allows to relate the properties of 
the states belonging to the doublets obtained combining $s_Q$ 
with the spin $s_\ell^P$ of the
light degrees of freedom.
The $SU(2 N_f)$ spin-flavour symmetry is explicitely 
broken by the finite values of the heavy quark masses, and the 
symmetry-breaking effects can be described using an 
expansion in the inverse masses of the heavy quarks \cite{hqet,hqet_rev}.

At the opposite energy scale, for vanishing masses of the up, down and strange 
quarks, QCD is invariant under chiral $SU(3)_L \times SU(3)_R$ transformations;
this symmetry is spontaneously broken, 
the Goldstone bosons being represented by
the octet of light pseudoscalar mesons \cite{chiral}. 

Both the heavy quark spin-flavour and the chiral symmetries can be
realized writing an effective QCD lagrangian,
expressed in terms of hadronic fields \cite{hqet_chir}. The
interactions of the heavy $s_\ell^P= {1 \over 2}^-$ and ${1 \over 2}^+$
mesons with the octet of light pseudoscalar $0^-$ mesons can be described 
by the lagrangian:
\begin{eqnarray}
{\cal L} &=& 
i\; Tr\{ H_b v^\mu D_{\mu ba} {\bar H}_a \}  + 
\frac{f_\pi^2}{8} Tr\{\partial^\mu\Sigma\partial_\mu \Sigma^\dagger \} 
+  Tr\{ S_b \;( i \; v^\mu D_{\mu ba} \; - \; \delta_{ba} \; \Delta) 
{\bar S}_a \} \nonumber\\
&+ & \; i \; 
g \; Tr\{H_b \gamma_\mu \gamma_5 {\cal A}^\mu_{ba} {\bar H}_a\} 
 +  i \; g' \; Tr\{S_b \gamma_\mu \gamma_5 {\cal A}^\mu_{ba} {\bar S}_a\}
+ \,[ i \, h \; Tr\{S_b \gamma_\mu \gamma_5 {\cal A}^\mu_{ba} {\bar H}_a\} \; 
+ \; h.c.] \;\;\; . \label{L}
\end{eqnarray}
The fields $H_a$ in (\ref{L}) 
describe the negative parity  $J^P=(0^-,1^-)$ ${\bar q }Q$ doublets 
(with $s_\ell^P= {1\over 2}^-$):
\begin{equation}
H_a = \frac{(1+{\rlap{v}/})}{2}[P_{a\mu}^*\gamma^\mu-P_a\gamma_5] \;\;,
\label{neg}
\end{equation}
where the operators 
$P^{*\mu}_a$ and $P_a$ respectively annihilate 
the $1^-$ ($B^*_a$) and $0^-$ ($B_a$) mesons of four-velocity $v$
($a=u,d,s$ is a light flavour index).
In the infinite heavy quark 
mass limit such states are degenerate in mass, 
due to the vanishing
of the chromomagnetic interaction of the heavy quark spin with the spin of the 
light degrees of freedom of the mesons.
Hence, $H_a$ describes a doublet of states with the light degrees of 
freedom with zero orbital angular momentum with respect to the heavy quark.
Analogously \cite{positivep}, 
the fields $S_a$ describe a doublet of states having 
the light degrees of freedom with angular momentum
$s_\ell={1\over 2}^+$ ($P$-waves in the constituent quark model):
\begin{equation}
S_a=\frac{1+{\rlap{v}/}}{2} \left[D_1^\mu\gamma_\mu\gamma_5-D_0\right] \;,
\label{pos}
\end{equation}
\noindent
with  $D_1^\mu$, $D_0$ annihilation operators for the states $1^+$ and $0^+$, 
respectively. Notice that all the heavy 
field operators in the lagrangian (\ref{L}) contain a factor $\sqrt{m_P}$
and have dimension $3/2$; the parameter $\Delta$ 
in (\ref{L}) represents the mass splitting between the 
positive and negative parity states. 

The octet of light pseudoscalar mesons is included 
in the effective lagrangian (\ref{L})
using the exponential representation: 
$\displaystyle \xi=e^{i {\cal M} \over f_\pi}$ and  $\Sigma=\xi^2$; 
the matrix ${\cal M}$ contains the $\pi, K$ and $\eta$ fields:
\begin{equation}
{\cal M}=
\left (\begin{array}{ccc} 
\sqrt{\frac{1}{2}}\pi^0+\sqrt{\frac{1}{6}}\eta & \pi^+ & K^+\nonumber\\
\pi^- & -\sqrt{\frac{1}{2}}\pi^0+\sqrt{\frac{1}{6}}\eta & K^0\\ 
K^- & {\bar K}^0 &-\sqrt{\frac{2}{3}}\eta
\end{array}\right )
\end{equation} 
with $f_{\pi}=132 \; MeV$. Finally, the operators $D$ and $\cal A$ in (\ref{L}) 
are given by:
\begin{eqnarray}
D_{\mu ba}&=&\delta_{ba}\partial_\mu+{\cal V}_{\mu ba}
=\delta_{ba}\partial_\mu+\frac{1}{2}\left(\xi^\dagger\partial_\mu \xi
+\xi\partial_\mu \xi^\dagger\right)_{ba}\\
{\cal A}_{\mu ba}&=&\frac{1}{2}\left(\xi^\dagger\partial_\mu \xi-\xi
\partial_\mu \xi^\dagger\right)_{ba} \; .
\end{eqnarray}
\par
The effective Lagrangian (\ref{L}) can be generalized to include the 
interactions of heavy $s_\ell^P={3\over 2}^+$ states, as well as the 
interaction with the octet of low-lying vector mesons
\cite{nardullirev}.

The strong interactions between heavy $H_a$, 
$S_a$ mesons and the light pseudoscalar mesons,
as described by (\ref{L}),  
are determined by three couplings: $g$, $h$ and
$g^\prime$.
The spin symmetry implies that  the couplings to light mesons 
(from now on we refer to pions) do not depend on the particular  member of 
a doublet; for example, the couplings $g_{B^* B^* \pi}$ and $g_{B^* B \pi}$,
defined respectively by the matrix elements
\begin{equation}
<{\bar B}^{*0}(p,\epsilon_2) \pi^-(q)|B^{*-}(p+q,\epsilon_1)>={2 i \over 
f_\pi}g_{B^*B^* \pi} \epsilon_{\rho \sigma \alpha \beta} \epsilon_1^\rho 
\epsilon_2^{* \sigma} p^\alpha q^\beta  \label{gbsbspi} 
\end{equation}
\noindent and
\begin{equation}
<{\bar B}^{0}(p) \pi^-(q)|B^{*-}(p+q,\epsilon)>={2 m_B \over f_\pi} 
g_{B^*B \pi} \epsilon^{\mu} q_\mu \hskip 3 pt  \label{gbsbpi} 
\end{equation}
\noindent
($\epsilon$ is the $B^*$ polarization vector),
coincide with the same coupling $g$ in the limit $m_b \to \infty$. Analogously,
the constant $h$ describes the coupling to pions of whatever a member of the 
$H_a$ doublet and a member of the $S_a$ one, whereas
the strong interaction of positive parity $S_a$ states with pions 
is governed by the coupling $g^\prime$.
 
Differences between the couplings of states belonging to the same doublet 
are of order $1/m_Q$.
Such differences play an important role in 
the heavy meson phenomenology; for example, the difference between the
pion-vector-vector $g_{B^* B^* \pi}$ and the pion-pseudoscalar-vector 
$g_{B^* B \pi}$ coupling constant appears in the $1/m_Q$ correcting terms to 
the effective lagrangian (\ref{L}) \cite{boyd}, and must be taken into account 
in the calculation of the heavy meson hyperfine splitting 
\cite{hyper,dibartolomeo95}. 

In refs.\cite{colangelo94,colangelo95,cinesi} the couplings 
$g$ and $h$ were computed  using QCD sum rules and the short-distance 
expansion \cite{qcdsr},
in the infinite heavy quark mass limit.
The short-distance expansion was also adopted to calculate the couplings
$g_{B^* B \pi}$, $g_{B^* B^* \pi}$  
and $g_{B_0 B \pi}$ for finite masses of the heavy quarks
\cite{eletski85,colangelo94,colangelo95,dibartolomeo95,dosch96}; 
comparing the results derived in the asymptotic regime and for finite 
masses,  significant information was obtained on the size of $1/m_Q$ 
corrections.

A different approach to compute the strong couplings was proposed in 
\cite{belyaev95} and adopted in \cite{colangelo95,aliev96}, 
using sum rules based on the 
expansion near the light-cone \cite{lcqcdsr0,lcqcdsr,braun1}
\footnote{ Applications of the light-cone sum rule method to  semileptonic 
heavy meson decays can be found in \cite{lcsem} and in references therein.
}.
The parameters $g_{B^* B \pi}$, $g_{B_0 B \pi}$ were determined by  
this technique for finite 
masses of the charm and bottom quarks, and  the couplings $g$ and $h$
were obtained 
extrapolating to $m_Q \to \infty$ the results for finite
$m_c$ and $m_b$. 

In this paper we derive light-cone QCD sum rules in the limit $m_Q \to \infty$.
This allows us to perform a systematic 
comparison between the results of light-cone and short-distance QCD sum rules,
which can shed light on the critical parameters, the 
advantages and the drawbacks of both the approaches. As we shall see, in the 
infinite heavy quark mass limit a remarkable agreement between 
the two approaches is obtained, supporting
our confidence on such methods for the calculation of parameters 
which are sensitive 
to the nonperturbative dynamics of QCD.  

The plan of the paper is as follows. 
In Sec. II we apply light-cone sum rules to calculate the coupling $g$ 
and $g_{B^* B^* \pi}$; the numerical analysis of this channel is 
presented in Sec. III, together with a comparison with the results obtained for 
$g_{B^* B \pi}$.
In Sec. IV and Sec. V we analyze the
couplings $g'$ and $h$, respectively; in particular, we consider
the process $B_1 \to B_0 \pi$.
In Sec. VI we draw our conclusions.

\section{ Light-cone QCD sum rules for 
${\lowercase{g}}_{B^*B^*\pi}$ and \lowercase{$g$}}

Let us first consider the finite heavy mass case, and compute the coupling
$g_{B^* B^* \pi}$ defined in eq.(\ref{gbsbspi}).
The correlator of two vector currents $J_\mu={\bar q} \gamma_\mu b$  
between the vacuum state and a pion external state with momentum $q$:
\begin{eqnarray}
F_{\mu \nu}(p,q) & = & i \int dx <\pi^-(q)| T[ {\bar d}(x)\gamma_\mu b(x), 
{\bar b}(0) \gamma_\nu u(0)] |0> e^{i p x} \nonumber \\
& = &  i \epsilon_{\mu \nu \alpha \beta} p^\alpha q^\beta \; F 
\label{corr}
\end{eqnarray}
corresponds to the diagram depicted in Fig.1.
Considering a  pion on mass shell: $q^2=m_\pi^2 $ (we set $m_\pi =0$), 
the invariant function $F$ depends on $p^2$ and $(p+q)^2$.

Following the same idea underlying the usual QCD sum rule method,
a hadronic representation  can be given to the correlator (\ref{corr})
in terms of the lowest lying 
resonance, the $B^*$ pole, of higher resonances and a continuum of states.
In a double dispersion relation for $F$ in the variables 
$p^2$ and $(p+q)^2$:
\begin{equation}
F(p^2, (p+q)^2) = \int ds ds' {\rho^{had} (s, s') \over (s-p^2) (s'- (p+q)^2)}
\;\;\;, \label{disprel}
\end{equation}
the spectral function $\rho^{had}$ gets contributions from $B^*$ and from 
higher states; therefore, it can be modeled as
\begin{equation} 
\rho^{had}(s,s')= {2 \over f_\pi} g_{B^*B^*\pi} m_{B^*}^2 f_{B^*}^2 
\delta(s-m_{B^*}^2) \delta(s'-m_{B^*}^2)
+ \rho^{cont}(s,s') \theta(s-s_0) \theta(s'-s'_0)\;\; \label{pole}
\end{equation}
where $\rho^{cont}(s,s')$ includes the contribution of the higher states and of 
the continuum above the effective thresholds $s_0, s'_0$;
the $B^*$ leptonic constant $f_{B^*}$  in (\ref{pole}) is defined as
\begin{equation}
<0|{\bar d} \gamma_\mu b |{\bar B}^{*0}(p,\epsilon)>=m_{B^*} f_{B^*} 
\epsilon_\mu \hskip 3 pt .\label{fb*} 
\end{equation}
We neglect in the dispersion relation (\ref{disprel}) possible subtraction 
terms, which play no role in the Borel transformed sum rules, as shown 
in the following.

For large positive values of $-p^2$ and $-(p+q)^2$
the function $F(p^2, (p+q)^2)$ can be computed in QCD.
This task is afforded expanding the T-product in eq.(\ref{corr}) 
near the light-cone $(x^2=0)$ in terms of non local operators, 
whose matrix elements are defined by wave functions associated with operators 
of increasing twist.
According to the notations in ref.\cite{belyaev95},
the catalog of the relevant matrix elements, up to twist 4, is: 
\begin{eqnarray}
<\pi(q)| {\bar d} (x) \gamma_{\mu} \gamma_5 u(0) |0>&=&-i f_{\pi} q_{\mu} 
\int_0^1 du \; e^{iuqx} (\varphi_{\pi}(u) +x^2 g_1(u) + {\cal O}(x^4) ) 
\nonumber \\
&+& f_\pi \big( x_\mu - {x^2 q_\mu \over q x} \big) 
\int_0^1 du \; e^{iuqx}  g_2(u) \hskip 3 pt  , \label{ax} \\
<\pi(q)| {\bar d} (x) i \gamma_5 u(0) |0> &=& {f_{\pi} m_{\pi}^2 \over m_u+m_d} 
\int_0^1 du \; e^{iuqx} \varphi_P(u)  \hskip 3 pt ,
 \label{pscal}  \\
<\pi(q)| {\bar d} (x) \sigma_{\mu \nu} \gamma_5 u(0) |0> &=&i(q_\mu x_\nu-q_\nu 
x_\mu)  {f_{\pi} m_{\pi}^2 \over 6 (m_u+m_d)} 
\int_0^1 du \; e^{iuqx} \varphi_\sigma(u)  \hskip 3 pt .
 \label{psigma}
\end{eqnarray}
\noindent 
The wave function $\varphi_{\pi}$ is associated with the leading twist 2 
operator, 
$g_1$ and $g_2$ correspond to twist 4 operators, and $\varphi_P$ and 
$\varphi_\sigma$ to twist 3 ones.
Due to the choice of the
light-cone gauge  $x^\mu A_\mu(x) =0$, 
the path-ordered gauge factor
$P \exp\big(i g_s \int_0^1 du x^\mu A_\mu(u x) \big)$ has been omitted.
Notice that the coefficient in front of the r.h.s. of 
eqs.(\ref{pscal}), (\ref{psigma})
can be written in terms of the light quark condensate
$<{\bar u} u>$ using the PCAC relation: 
$\displaystyle {f_\pi m_\pi^2 \over m_u+m_d}=-{2 \over f_\pi} <{\bar u} u>$.

The pion matrix elements of quark-gluon operators 
can also be parameterized in terms of wave functions \cite{belyaev95}:
\begin{eqnarray}
& &<\pi(q)| {\bar d} (x) \sigma_{\alpha \beta} \gamma_5 g_s 
G_{\mu \nu}(ux)u(0) |0>=
\nonumber \\
&&i f_{3 \pi}[(q_\mu q_\alpha g_{\nu \beta}-q_\nu q_\alpha g_{\mu \beta})
-(q_\mu q_\beta g_{\nu \alpha}-q_\nu q_\beta g_{\mu \alpha})]
\int {\cal D}\alpha_i \; 
\varphi_{3 \pi} (\alpha_i) e^{iqx(\alpha_1+v \alpha_3)} \;\;\; ,
\label{p3pi} 
\end{eqnarray}

\begin{eqnarray}
& &<\pi(q)| {\bar d} (x) \gamma_{\mu} \gamma_5 g_s 
G_{\alpha \beta}(vx)u(0) |0>=
\nonumber \\
&&f_{\pi} \Big[ q_{\beta} \Big( g_{\alpha \mu}-{x_{\alpha}q_{\mu} \over q \cdot 
x} \Big) -q_{\alpha} \Big( g_{\beta \mu}-{x_{\beta}q_{\mu} \over q \cdot x} 
\Big) \Big] \int {\cal{D}} \alpha_i \varphi_{\bot}(\alpha_i) 
e^{iqx(\alpha_1 +v \alpha_3)}\nonumber \\
&&+f_{\pi} {q_{\mu} \over q \cdot x } (q_{\alpha} x_{\beta}-q_{\beta} 
x_{\alpha}) \int {\cal{D}} \alpha_i \varphi_{\|} (\alpha_i) 
e^{iqx(\alpha_1 +v \alpha_3)} \hskip 3 pt  \label{gi} 
\end{eqnarray}
\noindent and
\begin{eqnarray}
& &<\pi(q)| {\bar d} (x) \gamma_{\mu}  g_s \tilde G_{\alpha \beta}(vx)u(0) |0>=
\nonumber \\
&&i f_{\pi} 
\Big[ q_{\beta} \Big( g_{\alpha \mu}-{x_{\alpha}q_{\mu} \over q \cdot 
x} \Big) -q_{\alpha} \Big( g_{\beta \mu}-{x_{\beta}q_{\mu} \over q \cdot x} 
\Big) \Big] \int {\cal{D}} \alpha_i \tilde \varphi_{\bot}(\alpha_i) 
e^{iqx(\alpha_1 +v \alpha_3)}\nonumber \\
&&+i f_{\pi} {q_{\mu} \over q \cdot x } (q_{\alpha} x_{\beta}-q_{\beta} 
x_{\alpha}) \int {\cal{D}} \alpha_i \tilde \varphi_{\|} (\alpha_i) 
e^{iqx(\alpha_1 +v \alpha_3)} \hskip 3 pt . \label{git} 
\end{eqnarray}
\noindent 
The operator $\tilde G_{\alpha \beta}$  is the dual of $G_{\alpha \beta}$:
$\tilde G_{\alpha \beta}= {1\over 2} \epsilon_{\alpha \beta \delta \rho} 
G^{\delta \rho} $; ${\cal{D}} \alpha_i$ is defined as 
${\cal{D}} \alpha_i =d \alpha_1 
d \alpha_2 d \alpha_3 \delta(1-\alpha_1 -\alpha_2 
-\alpha_3)$. 
The function $\varphi_{3 \pi}$ is twist 3, while all the wave 
functions appearing in eqs.(\ref{gi}), (\ref{git}) are twist 4.
The wave functions $\varphi (x_i,\mu)$ ($\mu$ is the renormalization point) 
describe the distribution in longitudinal momenta inside the pion, the 
parameters $x_i$ ($\sum_i x_i=1$) 
representing the fractions of the longitudinal momentum carried 
by the quark and the antiquark. 

The wave function normalizations immediately follow from the definitions
(\ref{ax})-(\ref{git}):
$\int_0^1 du \; \varphi_\pi(u)=\int_0^1 du \; \varphi_\sigma(u)=1$,
$\int_0^1 du \; g_1(u)={\delta^2/12}$,
$\int_0^1 du \; G_2(u)= {\delta^2/18}$
(with $G_2$ defined as $G_2(u)=-\int_0^u dv \; g_2(v)$)
$\int {\cal D} \alpha_i \varphi_\bot(\alpha_i)=
\int {\cal D} \alpha_i \varphi_{\|}(\alpha_i)=0$,
$\int {\cal D} \alpha_i \tilde \varphi_\bot(\alpha_i)=-
\int {\cal D} \alpha_i \tilde \varphi_{\|}(\alpha_i)={\delta^2/3}$,
with the parameter $\delta$ defined by the matrix element:
$<\pi(q)| {\bar d} g_s \tilde G_{\alpha \mu} \gamma^\alpha u |0>=
i \delta^2 f_\pi q_\mu$.

In the limit $\mu \to \infty$ the wave 
functions assume an asymptotic form which is approximately symmetric  in the 
variables $x_i$ and hence represents an almost equal distribution of the pion 
momentum between its constituents. This asymptotic wave function can be 
calculated by exploiting the conformal invariance of QCD in the 
short distance region. On the other hand, to investigate the general properties 
of the wave functions and their deviation from the asymptotic form, non 
perturbative methods are required. Usually, 
the momenta of the wave functions are determined by QCD sum rules, and then 
the wave functions themselves are reconstructed 
\cite{belyaev95,braun1}.
In a calculation of the type presented here, however, the complete form of the 
pion wave functions is not required, since only the values in particular 
points, namely near the symmetric point $u_0=1/2$, are needed. Therefore, the 
uncertainty in the results is only related to a finite number of input 
parameters, as it will be shown in the following.

The expansion of the correlator (\ref{corr}) near the light-cone produces the 
following  expression for the invariant function $F$:
\begin{eqnarray}
F^{QCD}(p^2,(p+q)^2)&=&f_\pi \int_0^1 du  {\varphi_{\pi}(u) \over 
m_b^2-(p+uq)^2} \nonumber \\
&-&f_\pi \int_0^1 du {4 \over [m_b^2-(p+uq)^2]^2} \Big[1 + {2 m_b^2 \over
m_b^2-(p+uq)^2} \Big] (g_1(u) + G_2(u))  \nonumber \\
&+&2 m_b {f_\pi m_\pi^2 \over 6 (m_u + m_d)}\int_0^1 du 
{\varphi_\sigma(u)\over [m_b^2-(p+uq)^2]^2} \nonumber\\
&+& f_\pi \int_0^1 dv \int {\cal D} \alpha_i {\varphi_\bot (\alpha_i)-
\tilde \varphi_\bot (\alpha_i) \over \{m_b^2-[p+q(\alpha_1+v \alpha_3)]^2\}^2} 
\label{fqcd} \\
&-& 4 f_\pi \int_0^1 dv \; (v-1) \int_0^1 d \alpha_3 {\hat \psi} (\alpha_3) 
{p \cdot q \over \{m_b^2-[p+q(1+(v-1) \alpha_3)]^2\}^3} \nonumber \\
&-& 4 f_\pi \int_0^1dv \int_0^1 d \alpha_3 \int_0^{1-\alpha_3} d \alpha_1 
{\hat \phi}(\alpha_1,\alpha_3) { p \cdot q \over 
\{m_b^2-[p+q(\alpha_1+v \alpha_3)]^2\}^3 } \nonumber \\
&+& 2 f_\pi \int_0^1 v \;  dv 
\int {\cal D} \alpha_i { \varphi_{\|}(\alpha_i) \over
\{m_b^2-[p+q(\alpha_1+v \alpha_3)]^2\}^2} \nonumber
\end{eqnarray}
\noindent where  ${\hat \psi}(\alpha_3)
=-\int_0^{\alpha_3} dt \int_0^{1-t} \phi(\alpha_1,t) d \alpha_1$, 
${\hat \phi}(\alpha_1,\alpha_3)=-\int_0^{\alpha_1} dt \phi(t,\alpha_3)$, and 
$\phi=\varphi_\bot-\tilde \varphi_\bot + 
\varphi_\| -\tilde \varphi_\|$.
Taking into account the wave function normalizations 
we recover, in the soft pion limit $q \to 0$, the expressions obtained in 
\cite{dibartolomeo95}. 

The sum rule for $g_{B^* B^* \pi}$ follows from the approximate equality of 
the expressions  (\ref{disprel}) and (\ref{fqcd}). Moreover, invoking duality 
arguments, the contribution of the continuum in (\ref{disprel}) corresponds to 
the QCD contribution. This allows to isolate the pole contribution to 
(\ref{disprel}). 
The equality between
(\ref{disprel}) and (\ref{fqcd}) can be improved
performing a double Borel transform in the 
variables $-p^2$ and $-(p+q)^2$. Defining 
$M_1^2$ and $M_2^2$ as the Borel parameters 
associated  to the channels $p$ and $p+q$, respectively, and using the formula:
\begin{equation}
{\cal B}_{M_1^2} {\cal B}_{M_2^2} 
{ (\ell-1)! \over [m_b^2 - (p + u q)^2]^\ell} =  
{ (M^2)^{2-\ell} \over M_1^2 M_2^2} \exp(- {m_b^2 + q^2(1-u_0) \over M^2}) \; 
\delta(u-u_0)
\end{equation}
with $\displaystyle M^2={M_1^2 M_2^2 \over M_1^2+M_2^2}$ and
$\displaystyle u_0={M_1^2  \over M_1^2+M_2^2}$, we get:
\begin{eqnarray}
F^{QCD}(M_1^2,M_2^2) &=&{f_\pi \over M_1^2 M_2^2}  
e^{-{m_b^2 \over M^2}} \Big\{  
M^2 \; \varphi_\pi (u_0) + 
{m_b  m_\pi^2 \over 3 (m_u+m_d)} \; \varphi_\sigma(u_0) \nonumber \\ 
&-&4  \Big(1+{m_b^2 \over M^2} \Big) 
[g_1(u_0)+G_2(u_0)] \nonumber \\
&+&  \int_0^{u_0} d \alpha_1 \int_{u_0-\alpha_1}^{1-\alpha_1} 
{d \alpha_3 \over \alpha_3} \; 
(\varphi_\bot -\tilde \varphi_\bot)\; (\alpha_1,1-\alpha_1-\alpha_3,
\alpha_3) \nonumber \\
&+& 2  
\int_0^{u_0}d \alpha_1 \int_{u_0-\alpha_1}^{1-\alpha_1} d \alpha_3 
\; {u_0-\alpha_1 \over \alpha_3^2} 
\; \varphi_\| (\alpha_1,1-\alpha_1-\alpha_3,
\alpha_3) \label{borelf} \\
&-&  \Big[{{\hat \psi} (1-u_0) \over 1-u_0} - \int_{1-u_0}^1 d \alpha_3 \;  
{{\hat \psi} (\alpha_3) \over \alpha_3^2} \Big] \nonumber \\
&-& \Big[ \int_0^{u_0} d \alpha_3  {{\hat \phi} (u_0-\alpha_3, \alpha_3) \over 
\alpha_3} -\int_0^1 d \alpha_3 {{\hat \phi} (u_0, \alpha_3) \over 
\alpha_3} \Big] \Big\} \;\;\; . \nonumber  
\end{eqnarray}
\noindent
On the other hand the Borel transformed hadronic representation of the function 
F reads:
\begin{equation}
F(M_1^2,M_2^2) = {2 \over f_\pi} 
{m^2_{B^*} f^2_{B^*} g_{B^* B^* \pi} \over  M_1^2 M_2^2} 
e^{-m^2_{B^*} ({1\over M_1^2} + {1\over M_2^2}) } 
+ \int ds ds' \rho^{cont}(s,s') e^{-{s \over M_1^2} - {s' \over M_2^2}} \;\;. 
\label{hadb} \end{equation}
As shown in (\ref{hadb}), 
the Borel transformation exponentially suppresses, for small values of the 
parameters $M_1^2, M_2^2$, 
the contribution of higher states and of the continuum; moreover, 
possible subtraction terms in (\ref{disprel}), which depend
 only on $p^2$ or $(p+q)^2$,
are removed by the independent borelization in the two channels. 
This is a feature which
renders light-cone sum rules quite appealing,
as far as the calculation of the strong couplings is concerned.
As a matter of fact, 
the determination of the strong couplings
by, e.g., short-distance rules is usually performed in the soft pion limit 
$q \to 0$, thus
preventing the possibility of an independent borelization in the two
$B^*$ channels. 

Considering the symmetry of the correlator (\ref{corr}) (Fig.1), 
it is natural to choose $M_1^2=M_2^2$, and then $u_0=1/2$, 
together with $s_0=s'_0$.
Such a choice corresponds
to a quark and an antiquark of the same momentum inside the pion. 
In this condition the subtraction of the continuum can be performed, 
in the leading twist  
term, with the substitution $e^{-m_b^2/M^2} \to 
e^{-m_b^2/M^2}-e^{-s_0/M^2} $
\cite{belyaev95}, a recipe we follow considering that the numerical 
contribution of the higher twist terms is small.

In this way we get the sum rule for $g_{B^* B^* \pi}$:
\begin{eqnarray}
f_{B^*}^2 g_{B^* B^* \pi} &=&{f_\pi^2 \over 2 m_{B^*}^2} 
e^{m_{B^*}^2 \over M^2}
\Bigg \{ (e^{-{m_b^2 \over M^2}}-e^{-{s_0\over M^2}}) 
M^2 \; \varphi_\pi (u_0) \nonumber \\
&+& e^{-{m_b^2 \over M^2}} \Big[  {m_b  m_\pi^2 \over 3 (m_u+m_d)} 
\; \varphi_\sigma(u_0)   
-4  \Big(1+{m_b^2 \over M^2} \Big) [g_1(u_0)+G_2(u_0)] \nonumber \\
&+&  \int_0^{u_0} d \alpha_1 \int_{u_0-\alpha_1}^{1-\alpha_1} 
{d \alpha_3 \over \alpha_3} \; 
(\varphi_\bot -\tilde \varphi_\bot)\; (\alpha_1,1-\alpha_1-\alpha_3,
\alpha_3) \nonumber \\
&+& 2  
\int_0^{u_0}d \alpha_1 \int_{u_0-\alpha_1}^{1-\alpha_1} d \alpha_3 
\; {u_0-\alpha_1 \over \alpha_3^2} 
\; \varphi_\| (\alpha_1,1-\alpha_1-\alpha_3,
\alpha_3) \label{sr} \\
&-&  \Big[{{\hat \psi} (1-u_0) \over 1-u_0} - \int_{1-u_0}^1 d \alpha_3 \;  
{{\hat \psi} (\alpha_3) \over \alpha_3^2} \Big] \nonumber \\
&-& \Big[ \int_0^{u_0} d \alpha_3  {{\hat \phi} (u_0-\alpha_3, \alpha_3) \over 
\alpha_3} -\int_0^1 d \alpha_3 {{\hat \phi} (u_0, \alpha_3) \over 
\alpha_3} \Big] \Bigg \} \; \;. \nonumber 
\end{eqnarray}

The numerical analysis of this sum rule is described in the next 
section. Here we want to show that eq.(\ref{sr}) has the right 
behaviour in the limit $m_b \to \infty$ \cite{belyaev95},
and that an analytical 
expression can be derived in the asymptotic regime.

In the 
limit $m_b \to \infty$ the $B^*$ meson mass can be related to the $b$ quark 
mass by the relation 
$m_{B^*}=m_b + \Lambda + {\cal O}({1 \over m_b })$, 
where $\Lambda$ represents the binding energy of the light degrees of freedom 
in the static $b$ quark chromomagnetic field. Moreover, the $B^*$ leptonic 
constant depends on $m_b$ as:
$f_{B^*}={{\hat F} \over \sqrt{m_b}}$
(modulo logarithmic corrections), with $\hat F$ a low energy parameter
remaining constant in the infinite $m_b$ limit.
Rescaling the Borel parameter: $M^2=2 m_b E$, and the 
continuum threshold  $s_0=m_b^2+ 2 m_b y_0$, with 
$E$ and $y_0$ independent of the heavy mass, we recover 
from (\ref{sr})  
a sum rule holding in the asymptotic limit:
\begin{equation}
{\hat F}^2 g = f_\pi^2  e^{\Lambda \over E}
\Bigg \{ \Big( 1- e^{-{y_0 \over E}} \Big) E \varphi_\pi(u_0) 
 -{<{\bar u} u> \over 3 f_\pi^2} \varphi_\sigma (u_0) 
-{1 \over E} [g_1(u_0)+ G_2(u_0)] \Bigg\} \;\;, \label{sumr_g} 
\end{equation}
which is the analogous of the sum rule derived for $g$  \cite{colangelo94}
by the short distance 
expansion.  Notice that the sum rule (\ref{sumr_g})
is not hampered by the problem of the subtraction of the so-called "parasitic" 
contributions, which must be carefully treated in the short-distance 
calculation \cite{colangelo94}; as a matter of fact, 
it results from two independent 
borelizations with parameters $E_1$ and $E_2$, with
$u_0= {E_1\over E_1+E_2}$ and
$E={E_1 E_2 \over E_1+E_2}$.

Finally, for $m_b \to \infty$, it is worth observing that the light-cone sum 
rule for  $g_{B^* B \pi}$ \cite{belyaev95} 
coincides with eq.(\ref{sumr_g}), as required by the heavy spin symmetry.

\section{Numerical analysis for ${\lowercase{g}}_{B^*B^*\pi}$ and
\lowercase{$g$}}

The numerical analysis of eqs. (\ref{sr}), (\ref{sumr_g})
 requires a set of light-cone
pion wave functions.
Actually, if the choice $\displaystyle u_0={1\over2}$ is made, 
only the value of the wave 
functions in the middle point is needed. 

Different sets of wave functions are available in the literature.
Here we use the functions derived in \cite{braun1} 
and adopted in \cite{belyaev95}
to compute the coupling $g_{B^* B \pi}$.
At the $b$-quark mass scale the values of the various functions appearing 
in eq.(\ref{sr}), at $u_0={1\over2}$, are: $\varphi_\pi(u_0)=1.219$,
$\varphi_\sigma(u_0)=1.463$,
$g_1(u_0)=-3.37 \times 10^{-3} \; GeV^2$ and
$G_2(u_0)=-1.8 \times 10^{-2} \; GeV^2$; these two last numbers correspond
to the choice $\delta^2(m_b)=0.17 \; GeV^2$.
The expressions for the functions $\phi$ and $\tilde \phi$ can be found in
\cite{belyaev95}; their contribution to the sum rule corresponds to a small 
fraction (of the order of $10^{-2}$) of the final result.

Using $m_b=4.6  \; GeV$ \cite{pich97}, $<{\bar u} u>=-(240 \; MeV)^3$
and the threshold $s_0$ in the range $s_0=33-35\; GeV^2$
we get the result depicted in Fig.2. The duality domain in $M^2$,
where the continuum contribution is small (less than $30 \; \%$),
can be chosen as
$M^2=6-12 \; GeV^2$. 
The result is:
\begin{equation}
f_{B^*}^2 \; g_{B^*B^*\pi} =(8.5 \pm 1.5) \times 10^{-3} \; GeV^2 \;\;\;,
\label{gb} \end{equation}
\noindent where the uncertainty is due to the variation with the threshold 
$s_0$ and the Borel parameter $M^2$.
We do not include the uncertainty on the wave functions, which we consider as 
external input parameters.

Eq.(\ref{sr}) can also be used to compute the coupling $g_{D^* D^* \pi}$.
Using $m_c=1.35 \; GeV$ and $s_0=6-7 \; GeV^2$, taking into account the 
variation of the wave functions with the renormalization scale, we derive the 
result in Fig.3. In the range $M^2=2-4 \; GeV^2$ we get:
\begin{equation}
f_{D^*}^2 \; g_{D^*D^*\pi} =(1.8 \pm 0.3)\times 10^{-2} \; GeV^2 \;\;.
\label{gd} \end{equation}
The results (\ref{gb}), (\ref{gd})
favourably compare with the outcome of short-distance QCD sum rules 
\cite{dibartolomeo95}:
$f_{B^*}^2 g_{B^*B^*\pi}=(9.4 \pm 1.8) \times 10^{-3} \; GeV^2$,
$f_{D^*}^2 g_{D^*D^*\pi}=(1.7 \pm 0.4)\times 10^{-2} \; GeV^2$.

The analogous couplings $B^* B \pi$, $D^* D \pi$, computed using the same set 
of parameters,  are:
\begin{eqnarray}
f_B f_{B^*} \; g_{B^*B\pi} &=&(7.9 \pm 1.0) \times 10^{-3} \; GeV^2 
\label{gb1} \\
f_D f_{D^*} \; g_{D^*D\pi} &=&(1.5 \pm 0.2)\times 10^{-2} \; GeV^2 \;\;.
\label{gd1} 
\end{eqnarray}
to be compared with the short-distance based calculation \cite{colangelo94}:
$f_B f_{B^*} \; g_{B^*B\pi} = (7.0 \pm 1.5) \times 10^{-3} \; GeV^2$ and 
$f_D f_{D^*} \; g_{D^*D\pi} = (1.2 \pm 0.3)\times 10^{-2} \; GeV^2$.

The analysis of the asymptotic sum rule (\ref{sumr_g})
allows to derive $g$ and,  by taking the logarithmic derivative 
in $1/E$, the parameter $\Lambda$. 
Using $y_0=1.2-1.4 \; GeV$ and $E=0.8-2.0 \; GeV$, we get, from the curves
depicted in Fig.4:
\begin{eqnarray}
\Lambda &=& 0.40 \pm 0.07 \; GeV \\
{\hat F}^2 g &=& 0.031 \pm 0.006 \; GeV^3 \;\;; \label{gas}
\end{eqnarray}
the analogous result obtained in \cite{colangelo94} is
${\hat F}^2 g = 0.035 \pm 0.008 \; GeV^3$.

The calculation of the couplings can be done, using eqs.(\ref{gb}-\ref{gd1}),
once the 
leptonic constants are known. Since the correlator (\ref{corr}) has been
computed at the leading order in $\alpha_s$, it is consistent to 
use the leptonic constants determined at the same order. 
Then, one can use
\cite{colangelo91}:
$f_{B^*}=190\pm 10 \; MeV$,
$f_{D^*}=220\pm 24 \; MeV$,
$f_{B}=150\pm20 \; MeV$,
$f_{D}=170\pm10 \; MeV$, with the results:
\begin{eqnarray}
g_{B^* B^* \pi}&=&  0.24 \pm 0.05 \;\;\;, \hskip 1cm
g_{D^* D^* \pi}=  0.31 \pm 0.08 \;\; \\
g_{B^* B \pi}&=&  0.28 \pm 0.05 \;\;\;, \hskip 1cm
g_{D^* D \pi}=  0.40 \pm 0.07 \;\; .
\end{eqnarray}
To derive the value of $g$, in the limit $m_b \to \infty$,  one can linearly
extrapolate the above results: 
\begin{eqnarray}
g_{P^* P^* \pi}&=&g \; \big(1+{a \over m_Q}\big) \\
g_{P^* P \pi}&=&g \; \big(1+{b \over m_Q}\big) \;\; ,
\label{scale} 
\end{eqnarray}
obtaining:
\begin{eqnarray}
g=0.21\pm 0.07 \;\;\;\;\;\; &,& \;\;\;\;\;\; a=0.6\pm0.9 \; GeV \label{resg1} \\
g=0.23\pm 0.08 \;\;\;\;\;\; &,& \;\;\;\;\;\; b=1.0\pm1.0 \; GeV \label{resg2}
\end{eqnarray}
from $g_{P^*P^*\pi}$ and $g_{P^*P\pi}$, respectively.
The results (\ref{resg1}), (\ref{resg2}) show that the same value for $g$ is 
obtained by extrapolating to $m_b \to \infty$ the values obtained for
$g_{P^*P^*\pi}$ and $g_{P^*P\pi}$. One can conservatively quote
$g=0.22 \pm 0.10$ as the result of the extrapolation. It is also worth 
observing that, although the fit does not accurately fix the parameters $a,b$,
the $1/m_Q$ corrections are rather sizable in the case of the charm quark.

Let us now consider the calculation of $g$ from eq.(\ref{gas}).
Using the determination in \cite{neubertf}:
$\hat F=0.30 \pm 0.05$, obtained neglecting radiative ${\cal O}(\alpha_s)$
corrections, one gets:
\begin{equation}
g=0.34 \pm 0.10 \;\;\; ,
\end{equation}
a result which, although higher in the central value, is compatible with the 
outcome of the fit.

It is interesting to consider the inclusion of 
$O(\alpha_s)$ corrections in the leptonic constants, 
to investigate at least
partially the role of the perturbative corrections in the determination of $g$.
Using, therefore, 
$f_{B^*}=213\pm 34 \; MeV$,
$f_{D^*}=258\pm 26 \; MeV$,
$f_{B}=180\pm30 \; MeV$,
$f_{D}=195\pm20 \; MeV$ \cite{colangelo91}, one gets:
\begin{eqnarray}
g_{B^* B^* \pi}&=&  0.19 \pm 0.04 \;\;, \hspace*{1cm}
g_{D^* D^* \pi}=  0.23 \pm 0.05 \\ 
g_{B^* B \pi}&=&  0.21 \pm 0.06 \;\;, \hspace*{1cm}
g_{D^* D \pi}=  0.30 \pm 0.06 \;\;,  
\end{eqnarray}
i.e.:
\begin{eqnarray}
g=0.17\pm 0.06 \;\;\;\;\;\; &,& \;\;\;\;\;\; a=0.6\pm0.8 \; GeV \\
g=0.17\pm 0.08 \;\;\;\;\;\; &,& \;\;\;\;\;\; b=1.0\pm1.4 \; GeV \;\;\;.
\end{eqnarray}
The inclusion of the ${\cal O}(\alpha_s)$ terms in the leptonic constants does 
not spoil the prediction that the same $g$ is obtained from the $B^*B^*\pi$ and 
$B^* B \pi$ channels. The numerical value of $g$ results modified at the level 
of $30 \%$.

Several determinations of the coupling $g$ can be found in the literature 
\cite{colangelo94,dibartolomeo95,dosch96,belyaev95,noipot,altre}, 
with values varying up to $g=1$. The analysis reported here points
towards small values of $g$, in agreement with the outcome of the
relativistic potential model \cite{noipot}, which shows that
the nonrelativistic result $g=1$ is reduced ($g \simeq 1/3$) if a 
relativistic treatment of the light quarks is adopted.

\section{Light-cone sum rules for $\lowercase{g}_{B_1 B_0 \pi}$ and 
$\lowercase{g'}$}
The same method described in the previous sections can be used to determine
the coupling $g^\prime$, which weightes  in the lagrangian (\ref{L})
the strong interaction between positive parity heavy mesons and pions.
For a finite mass of the heavy quark,
the coupling $g_{B_1 B_0 \pi}$ can be obtained from the correlator of 
a scalar $J_S=\bar q Q$
and an axial $J_a^\mu=\bar q \gamma^\mu \gamma_5 Q$ current:
\begin{eqnarray}
G_\mu (p,p+q)&=&i \int dx <\pi^-(q)|T[{\bar d}(x) b(x), 
{\bar b} (0)  \gamma_\mu \gamma_5 u(0)]|0> e^{ipx} \nonumber \\
&=&G^{(0)} (p+q)_\mu + G^{(1)} q_\mu \;\;\; \label{decg}
\end{eqnarray}
\noindent 
since the invariant function $G^{(1)}$
gets contribution from the poles $B_0$ and $B_1$. 
Defining
\begin{equation}
<\pi(q) B_0(p)|B_1(p+q, \epsilon)>= { 2 \sqrt{m_{B_0} m_{B_1}}\over f_\pi}
g_{B_1 B_0 \pi} \; \epsilon \cdot q
\hskip 3 pt  \label{gb0} 
\end{equation}
\noindent
with $m_{B_1}$ and $m_{B_0}$ the masses of the positive $1^+$ and $0^+$ states, 
respectively, and 
\begin{equation}
<0|{\bar b} u|B_0> ={f_{B_0} m_{B_0}^2 \over m_b} \;\;\;, \label{fb0} 
\end{equation}
\noindent 
\begin{equation}
<0|{\bar d} \gamma_\mu \gamma_5 b|B_1(p,\epsilon)>=f_{B_1} m_{B_1} 
\epsilon_\mu \;\;,
\label{fb1} \end{equation}
we get the Borel transformed sum rule for $g_{B_1 B_0 \pi}$:
\begin{eqnarray}
g_{B_1 B_0 \pi}&& 
{2 f_{B_1} m_{B_1}^{3\over2} f_{B_0} m_{B_0}^{5\over2} \over f_\pi m_b}=
e^{m_{B_0}^2+m_{B_1}^2 \over 2 M^2} \Bigg\{
(e^{-m_b^2/M^2}-e^{-s_0/M^2}) M^2 
 m_b f_\pi \varphi_\pi(u_0) \nonumber \\
&+& e^{-m_b^2/M^2}
\Bigg[ M^2\Big[-{f_\pi m_\pi^2 \over m_u+m_d}(1-u_0) \varphi_P(u_0)- 
{f_\pi m_\pi^2 \over 6 (m_u+m_d)} 
\Big(2 \varphi_\sigma(u_0)-(1-u_0) {d \varphi_\sigma \over du} (u_0) 
\Big) \Big] \nonumber \\
& -&\Big[2 m_b f_\pi (1-u_0)g_2(u_0)+{f_\pi m_\pi^2 \over 6 (m_u+m_d)}2 m_b^2 
\varphi_\sigma(u_0) \Big] \nonumber \\
&-& {1 \over M^2}4 m_b^3 f_\pi [g_1(u_0)+G_2(u_0)] \label{g10pi} \\
&-&m_b f_\pi \int_0^{u_0} d \alpha_1 \int_{u_0-\alpha_1}^{1-\alpha_1} 
d \alpha_3 
\; (-\varphi_\| +\tilde \varphi_\|+2\varphi_\bot -2 \tilde \varphi_\bot) 
(\alpha_1, 1-\alpha_1-\alpha_3, \alpha_3) \nonumber \\
&-&2 M^2 f_{3 \pi} 
\Big[ \int_0^{1-u_0} {d \alpha_3 \over \alpha_3} \varphi_{3 \pi} 
(u_0, 1-u_0-\alpha_3, \alpha_3) \nonumber \\
&-& \int_0^{u_0} d \alpha_1 
\int_{u_0-\alpha_1}^{1-\alpha_1} {d \alpha_3 \over \alpha_3^2}\varphi_{3 \pi}
(\alpha_1,1-\alpha_1-\alpha_3, \alpha_3) \Big] \Bigg] \Bigg\} \;\;\;. 
\nonumber 
\end{eqnarray}
\noindent 
In the infinite heavy quark mass limit, it is easy to derive 
from (\ref{g10pi}) the sum rule for $g^\prime$,
using the scaling law 
$f_{B_1}={\hat F^+ \over \sqrt{m_b}}$:
\begin{equation}
\hat F^{+2} g^\prime =  {f_\pi \over 2}  e^{\Lambda^\prime \over E}
\Bigg \{ \Big( 1- e^{-{y_0 \over E}} \Big)  2 E f_\pi \varphi_\pi(u_0) + 
 {2 \over 3} {<{\bar u} u> \over  f_\pi} \varphi_\sigma (u_0) 
-{2 f_\pi \over E} [g_1(u_0) + G_2(u_0)] \Bigg\}  \;\;\;. \label{sumr_gp} 
\end{equation}
\noindent 
$E$ is the Borel parameter; 
$\Lambda^\prime$ is the binding energy for the $s_\ell={1\over2}^+$ doublet:
$\Lambda^\prime=m_{B_{1,0}}-m_b$, and it can be obtained by the logarithmic 
derivative of (\ref{sumr_gp}).

Before performing the numerical analysis of  eq.(\ref{g10pi}), we notice
a potential numerical problem related to the presence 
of terms of opposite signs and similar sizes contributing to the sum rule.
An example is represented by the first two terms  in (\ref{g10pi}), 
respectively proportional to $\varphi_\pi$ and $\varphi_P$. 
For intermediate values of the heavy quark masses
there is no evidence of a hierarchical structure 
among the contributions from different twists;
moreover, cancellations between various terms results in
a critical dependence on the values of the wave functions, which have
their own uncertainties.
On the other hand, in the infinite heavy quark mass limit 
such a problem does not occur, and therefore eq.(\ref{sumr_gp}) provides us 
with the possibility of determining $g^\prime$ more reliaby.

Using 
$m_{B_1}=m_{B_0}=5.7\; GeV$ and $s_0=36-38 \; GeV^2$ we get
\begin{equation}
f_{B_1} f_{B_0} \; g_{B_1 B_0 \pi} =(1.2 \pm 0.4) \times 10^{-3} \; GeV^2 
\;\; . \label{res_g1}
\end{equation}
A determination of $g_{P_1 P_0 \pi}$ immediately follows, using the leptonic
constants of  $B_1$ and $B_0$ computed in 
\cite{colangelo91}:
$f_{B_1}=180\pm 30 \; MeV$,
$f_{B_0}=180\pm 30 \; MeV$
(the corresponding constants in
the charm case are  $f_{D_1}=240\pm20 \; MeV$ and $f_{D_0}=170\pm20 \; MeV$).
We get:
\begin{equation}
g_{B_1 B_0 \pi} =  0.05 \pm 0.02 
\;\; \;. \label{res_g1a} 
\end{equation}
Going to smaller heavy quark masses, the coupling
$g_{P_1 P_0 \pi}$ decreases. However, the cancellation between
different terms becomes more and more important, and the result
becomes critically dependent on the values of the wave functions.  

The asymptotic sum rule (\ref{sumr_gp}), 
depicted in Fig.5 and obtained 
varying $y_0$ in the range
$y_0=1.3-1.5 \; GeV$,  and the Borel parameter $E$ in the range 
$E=1.0-2.0 \; GeV$, provides the result:
\begin{eqnarray}
\Lambda^\prime &=& 0.89 \pm 0.05 \; GeV \\
\hat F^{+2} g^\prime &=& 0.022 \pm 0.003 \; GeV^3 \;\;\;. 
\end{eqnarray}
Using $\hat F^+=0.46\pm0.06 \; GeV^{3\over 2}$
\cite{colangelo91} the last equation corresponds to:
\begin{equation}
g^\prime = 0.10 \pm 0.02 \;\;\;. \label{res_gprime}
\end{equation}
 
The comparison between (\ref{res_g1a}) and (\ref{res_gprime})
gives an insight on the role of the heavy quark mass corrections in the case
of the determination of the strong coupling $g'$.

\section{ Light-cone QCD sum rule for 
${\lowercase{g}}_{B_1 B^*\pi}$ and $\lowercase{h}$}

To complete our analysis of the couplings appearing in (\ref{L}) 
we consider the parameter $h$ corresponding to the couplings
$B_1 B^* \pi$ and $B_0 B \pi$. 
The coupling $g_{B_0 B \pi}$ has been computed by light-cone 
and short-distance QCD sum rules  in \cite{colangelo95}, 
while a determination of $g_{B_1 B^* \pi}$ can be found in 
\cite{aliev96}. Here we want to derive an expression 
for $h$. 

Let us define $g_{B_1 B^* \pi}$ by the matrix element
\begin{equation}
<B^*(p,\epsilon) \pi(q) | B_1(p+q, \eta)>=g_{B_1 B^* \pi} (\epsilon^* \cdot 
\eta) {p \cdot q \over m_{B_1}} + (p \cdot \eta) (q \cdot \epsilon^*) F \;\;; 
\label{mgb1} 
\end{equation}
\noindent 
$g_{B_1 B^* \pi}$ is related to the coupling $h$ in eq.(\ref{L}) by:
\begin{equation}
g_{B_1 B^* \pi}=-{2 h \over f_\pi}  \sqrt{m_{B_1} m_{B^*}} \;\;\;, 
\label{rel} 
\end{equation}
whereas the structure $F$ is subleading in the $m_Q \to \infty$ limit.
To calculate $g_{B_1 B^* \pi}$ 
one has to consider the correlator of an axial and a vector current:
\begin{eqnarray}
H_{\mu \nu}(p,p+q)&=&
i \int dx <\pi^-(q)|T[{\bar d}(x) \gamma_\mu b(x), 
{\bar b} (0) \gamma_\nu \gamma_5 u(0)]|0> e^{ipx} \nonumber \\
& = & H^{(0)} g_{\mu \nu} + H^{(1)} p_\mu p_\nu + ...
\hskip 1 cm .\label{dec}
\end{eqnarray}
\noindent 
The $1^+$ and $1^-$ poles contribute to the 
invariant function $H^{(0)}$, which must be considered to derive 
$g_{B_1 B^* \pi}$. One obtains the Borel transformed sum rule:
\begin{eqnarray}
g_{B_1 B^* \pi} && (m_{B_1} - m_{B^*}) f_{B_1} m_{B_1}
 f_{B^*} m_{B^*} e^{-{m_{B^*}^2+m_{B_1}^2 \over 2 M^2}}= \nonumber \\
&=& f_\pi \Bigg\{ \Big( e^{-{m_b^2 \over M^2}}-e^{-{s_0 \over M^2}} \Big) 
\Big({M^4 \over 2} \varphi_\pi^\prime (u_0) +
{m_b m_\pi^2 \over m_u+m_d} M^2 \varphi_P (u_0) \Big) 
- e^{-m_b^2/M^2}\Bigg[ 2 m_b^2 g_2(u_0) \nonumber \\
&+&  2 M^2 
\Big(1+{m_b^2 \over M^2} \Big) [ G_2^\prime (u_0)+g_1^\prime(u_0)] \label{hsr}
\\
&+&M^2 \Big[ \int_0^{u_0} {d \alpha_3 \over 2 \alpha_3} (\varphi_\|+ 
\tilde \varphi_\|)(u_0-\alpha_3, 1 -u_0,\alpha_3) +
\int_0^1 {d \alpha_3 \over 2 \alpha_3} (\varphi_\|- 
\tilde \varphi_\|)(u_0, 1 -u_0-\alpha_3,\alpha_3) \nonumber \\
&-& \int_0^{u_0} d \alpha_1 \int_{u_0-\alpha_1}^{1-\alpha_1} {d \alpha_3 \over 
\alpha_3^2} \varphi_\|(\alpha_1, 1 -\alpha_1-\alpha_3, \alpha_3) \Big] \Bigg]
\Bigg\} \;\;\;.
\nonumber \end{eqnarray}

Differently from the calculation of $g$ and $g'$, 
the determination of $h$ does not 
concern a symmetric correlator (Fig.1). However, also in this case the choice 
$u_0=1/2$ allows us to use the same values of the pion wave functions employed 
for $g,g'$, which reduces the uncertainty related to the wave functions. 
Considering that
$\varphi_\pi^\prime \Big({1 \over 2}\Big)=
g_2\Big({1 \over 2}\Big)=G_2^\prime\Big({1 \over 2}\Big)=
g_1^\prime\Big({1 \over 2}\Big)=0$, in the
limit $m_b \to \infty $ one obtains from eq.(\ref{hsr}) the expression:
\begin{equation}
{\hat F}^+ \; {\hat F} \; h =
-{f_\pi^2 m_\pi^2 \over m_u + m_d} {E \over \Lambda^\prime 
-\Lambda} (1 - e^{-y_0/E}) e^{\Lambda^\prime+\Lambda \over 2 E} 
\varphi_P (u_0)\;\;\;. \label{hlim}
\end{equation}
Eq.(\ref{hlim}) coincides with the expression that can be 
obtained from the sum rule for
$g_{B_0 B \pi}$ \cite{colangelo95}.

The numerical results from eq. ({\ref{hsr}), using the same set of parameters 
chosen in the previous Sections:
\begin{equation}
f_{B_1} f_{B^*} g_{B_1 B^* \pi} =  1.9 \pm 0.4 \; GeV^2 \;\;, \hspace*{1cm}
f_{D_1} f_{D^*} g_{D_1 D^* \pi} =  1.2 \pm 0.2 \; GeV^2
\end{equation}
produce the following values for  $g_{P_1 P^* \pi}$:
\begin{equation}
g_{B_1 B^* \pi}=  56 \pm 15 \;\;, \hspace*{1cm}  
g_{D_1 D^* \pi}=  23 \pm 5  \;\;. 
\end{equation}
The linear extrapolation to $m_b\to\infty$:
$h_{P_1 P^* \pi}= h (1 +{d \over m_Q})$ determines the coupling $h$:
\begin{equation}
h= - 0.7 \pm 0.3\;\; \;\;, \hspace*{1cm}  d= 0.0 \pm 0.7  \; GeV \;\;\;.  
\end{equation}
The same result comes from 
eq.(\ref{hlim}); as a matter of fact, choosing 
$y_0=1.3-1.5 \; GeV$ and the mass parameters
$\Lambda, \; \Lambda'$ derived in the previous sections, the result
(Fig.6):
\begin{equation}
\hat F \; \hat F^+ \; h = - 0.083 \pm 0.005 \; GeV^3
\end{equation}
corresponds to
\begin{equation}
 h = - 0.60 \pm 0.13 \;\;, \label{has} 
\end{equation}
in agreement with the determination $h=-0.56 \pm 0.28$ obtained in 
\cite{colangelo95} from $g_{B_0B\pi}$.

\section{Conclusions}

The strong couplings of heavy mesons to pions are expected to obey symmetry 
relations in the $m_Q \to \infty$ limit, when QCD displays invariance under 
rotations in the heavy quark spin and flavour space. In this limit, it can be 
shown that the number of independent couplings reduces to a set of effective 
parameters. 

We employed light-cone sum rules to evaluate the couplings $g_{P^* P^* \pi}$,
$g_{P^* P_1 \pi}$, $g_{P_1 P_0 \pi}$, $(P=D,B)$ both for finite values of the 
heavy quark mass and in the infinite mass limit. This procedure allowed us 
to observe that the light-cone sum rule method reproduces the right 
$m_Q$ scaling 
behaviour for the physical quantities.
Moreover, we verified that the asymptotic results agree with usual QCD sum 
rules based on the short distance expansion. 
 This comparison enforces our confidence in the reliability of both the 
variants of the QCD sum rule tecnique (e.g. short distance based and light cone 
sum rules), which are among the few available instruments to investigate non 
perturbative aspects of QCD.

\vspace*{2cm}
\noindent{\bf Acknowledgments\\}
\noindent We thank G.Nardulli and N.Paver for interesting discussions.


\clearpage

\hskip 5 cm {\bf FIGURE CAPTIONS}
\vskip 1 cm

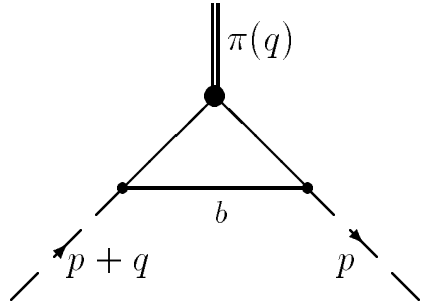
\noindent {\bf Fig. 1}\\
\noindent
The correlator in eq.(\ref{corr}).
\vspace{4mm}

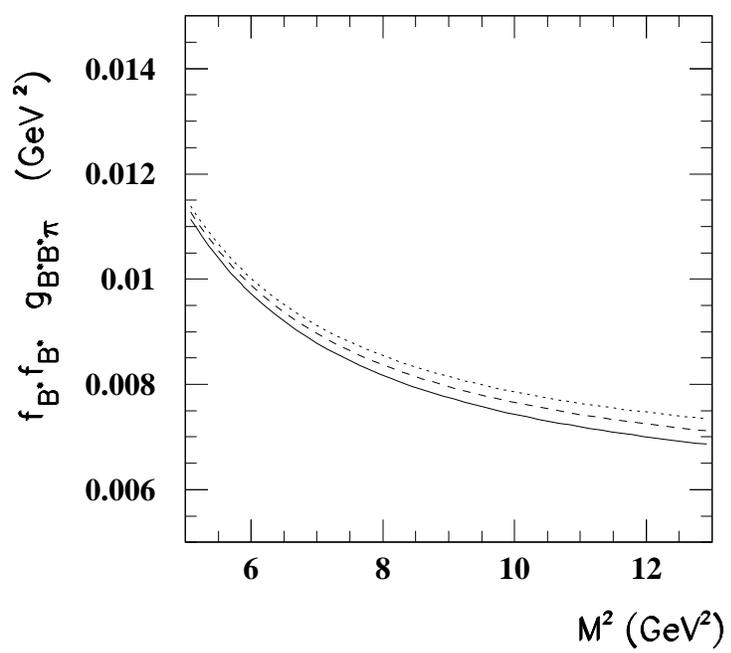
\noindent {\bf Fig. 2}\\
\noindent
Stability curve for $f_{B^*}^2 \; g_{B^*B^*\pi}$.
The curves correspond to the threshold $s_0=34 \; GeV^2$ (continuous line),
$s_0=35 \; GeV^2$ (dashed line), $s_0=36 \; GeV^2$ (dotted line).
\vspace{4mm}

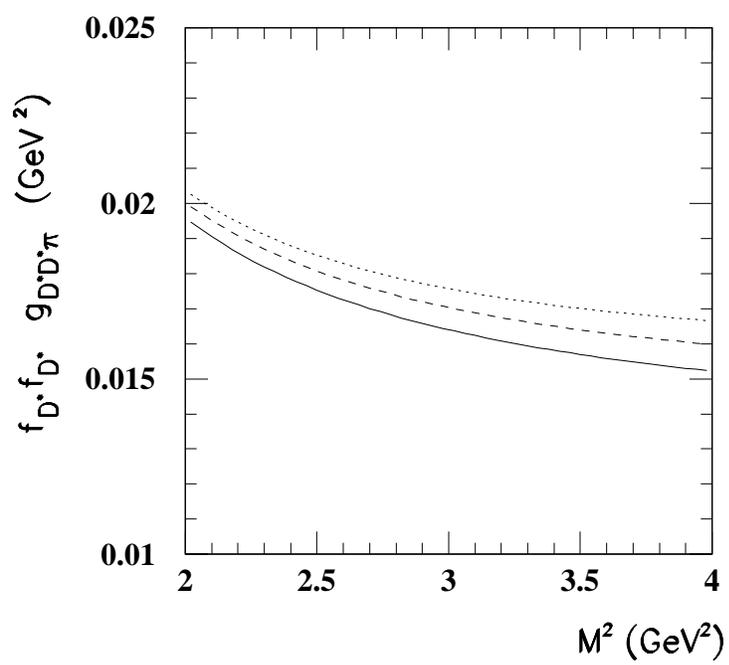
\noindent {\bf Fig. 3}\\
\noindent
Stability curve for $f_{D^*}^2 \; g_{D^*D^*\pi}$.
The curves correspond to $s_0=6 \; GeV^2$ (continuous line),
$s_0=6.5 \; GeV^2$ (dashed line), $s_0=7 \; GeV^2$ (dotted line).
\vspace{4mm}

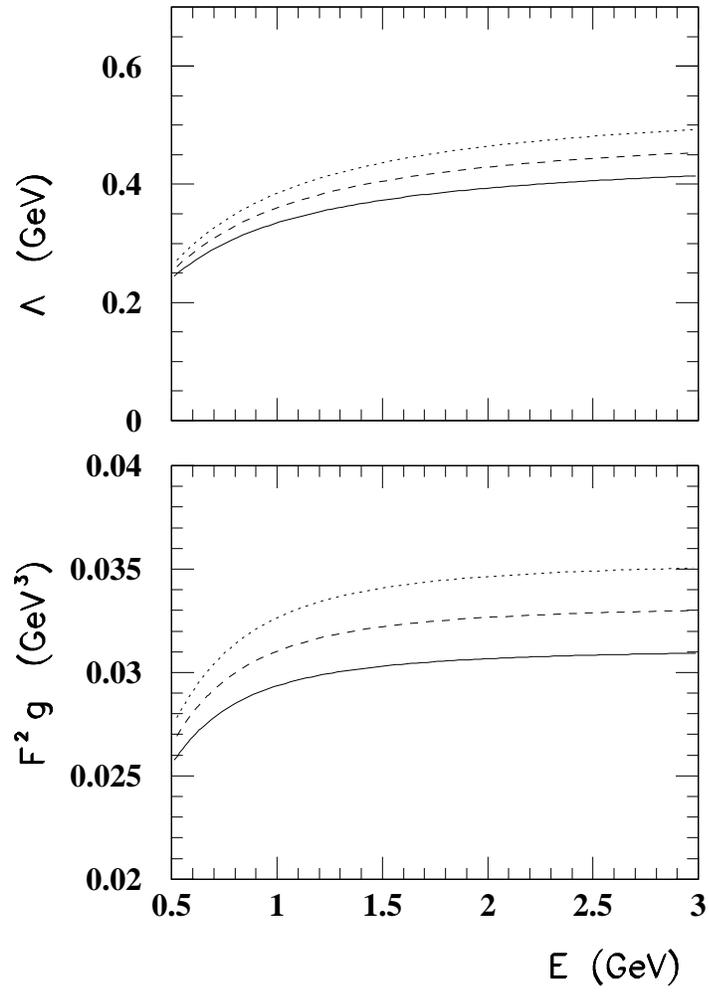
\noindent {\bf Fig. 4}\\
\noindent
Stability curve for $\Lambda$ and for $\hat F^2 g$.
The curves correspond to the threshold $y_0=1.2 \; GeV$ (continuous line),
$y_0=1.3 \; GeV$ (dashed line), $y_0=1.4 \; GeV$ (dotted line).
\vspace{4mm}

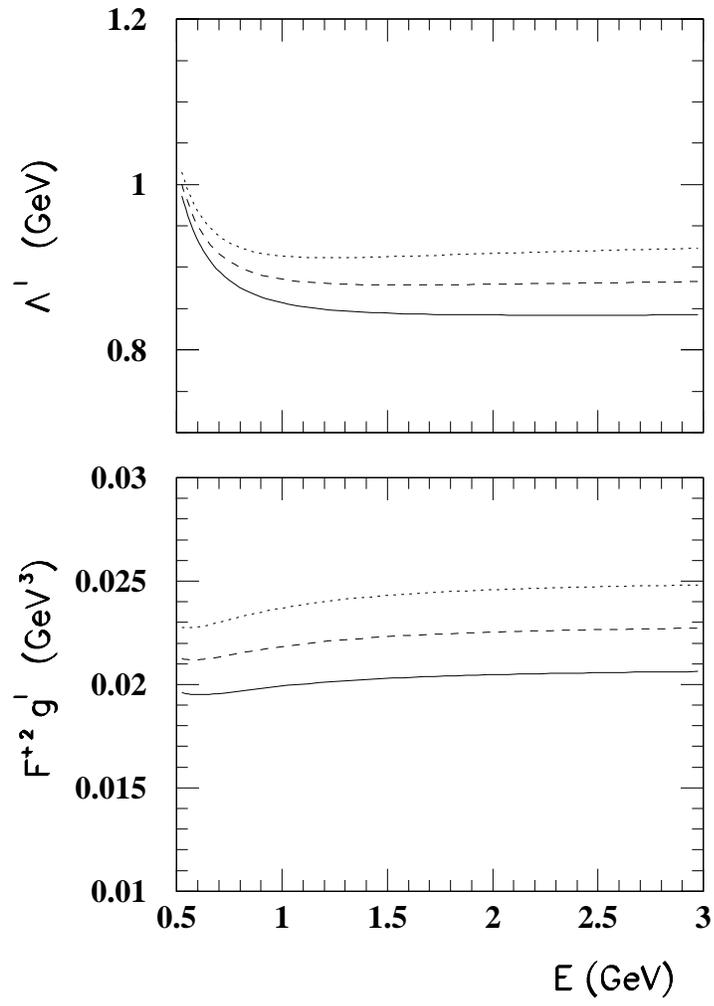
\noindent {\bf Fig. 5}\\
\noindent
Stability curve for $\Lambda'$ and for $\hat F^{+2} g'$.
The curves correspond to $y_0=1.3 \; GeV$ (continuous line),
$y_0=1.4 \; GeV$ (dashed line), $y_0=1.5 \; GeV$ (dotted line).
\vspace{4mm}

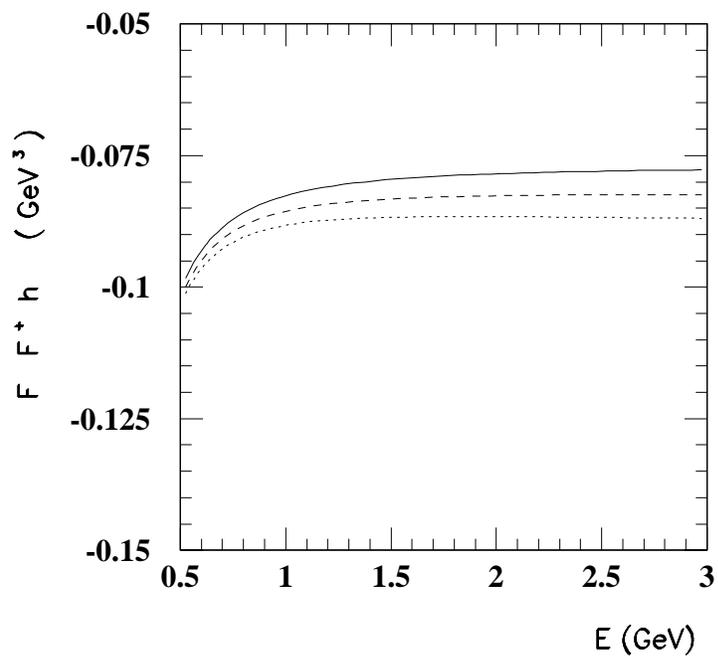
\noindent {\bf Fig. 6}\\
\noindent
Stability curve for $\hat F \hat F^+ h$.
The curves correspond to $y_0=1.3 \; GeV$ (continuous line),
$y_0=1.4 \; GeV$ (dashed line), $y_0=1.5 \; GeV$ (dotted line).
\vspace{4mm}

\end{document}